\documentclass[11pt,letterpaper]{article}

\usepackage[utf8]{inputenc}
\usepackage[T1]{fontenc}
\usepackage{times}
\usepackage{amsmath,amssymb,amsfonts}
\usepackage{graphicx}
\usepackage{float}
\usepackage{caption}
\usepackage{subcaption}
\usepackage{algorithm}
\usepackage{algpseudocode}
\usepackage[square, numbers, comma, sort&compress]{natbib}
\usepackage[hidelinks]{hyperref}
\usepackage{listings}
\usepackage{xcolor}
\usepackage{threeparttable}
\usepackage{booktabs}
\usepackage{multirow}
\usepackage[margin=1in]{geometry}
\usepackage{colortbl}

\graphicspath{{Figures/}}

\title{\textbf{Super-Resolution Enhancement of Medical Images Based on Diffusion Model: An Optimization Scheme for Low-Resolution Gastric Images}}

\author{
    Haozhe Jia\thanks{Department of Computer Science, Boston University, Boston, MA. Email: jimmyjia@bu.edu} \\
    \textit{Supervised by Subrota Kumar Mondal}
}

\date{}

\begin{document}

\maketitle


\begin{abstract}
The advent of capsule endoscopy has revolutionized gastrointestinal imaging by enabling minimally invasive internal visualization. However, its clinical utility is constrained by the inherently low resolution of the acquired imagery, resulting from limitations in onboard hardware, power supply, and wireless data transmission. These constraints impede the accurate identification of fine-grained mucosal textures, pathological features such as polyps or ulcerations, and subtle morphological variations essential for early diagnosis. This study addresses these challenges by investigating the efficacy of a diffusion-based super-resolution framework that can enhance capsule endoscopy images in a data-driven and anatomically consistent manner.

This research adopts the SR3 (Super-Resolution via Repeated Refinement) framework~\cite{saharia2021image}, built upon Denoising Diffusion Probabilistic Models (DDPMs), to learn a probabilistic mapping from low-resolution to high-resolution images. Unlike adversarial learning-based approaches that suffer from instability and hallucination artifacts, diffusion models offer robust likelihood-based training with reliable convergence behavior. The HyperKvasir dataset~\cite{Borgli2020}, a large-scale publicly available collection of gastrointestinal endoscopy images labeled by anatomical and pathological attributes, serves as the primary training and evaluation corpus. The model takes as input a six-channel concatenation of a bicubic-upsampled low-resolution image and random Gaussian noise, and progressively denoises it across 2000 diffusion steps to approximate the high-resolution ground truth. The network architecture is a U-Net backbone with hierarchical feature extraction, multiscale supervision, and group normalization.

Empirical evaluations demonstrate that the SR3 model significantly enhances image fidelity over traditional interpolation methods and GAN-based super-resolution frameworks such as ESRGAN. Quantitatively, the proposed approach achieves mean PSNR of 27.5 dB and SSIM of 0.65 for the first-generation model, improving to 29.3 dB and 0.71 for the second-generation model with attention mechanisms. Qualitatively, the model preserves mucosal boundaries, vascular patterns, and lesion structures with high anatomical faithfulness, which are critical for downstream tasks such as computer-aided detection and classification. The findings suggest that diffusion-based super-resolution techniques hold strong promise for advancing non-invasive medical imaging, particularly in capsule endoscopy where hardware-imposed constraints limit image resolution.

\end{abstract}


\section{Introduction}
\label{sec:intro}

Capsule endoscopy has emerged as a widely adopted non-invasive diagnostic tool for visualizing the gastrointestinal (GI) tract~\cite{Akpunonu200, RUEDA2013113}. However, due to the constraints of miniaturized camera sensors and energy efficiency, the resolution of images captured by capsule endoscopes remains relatively low. This limitation hinders the accurate identification of subtle pathological features, such as small polyps, ulcers, or bleeding points, which are essential for early diagnosis and effective treatment.

In contrast, traditional invasive endoscopy offers high-resolution imagery but at the cost of patient discomfort, procedural risk, and limited accessibility. Therefore, a significant research challenge lies in enhancing the resolution of non-invasive capsule images without compromising patient comfort. This forms the core motivation of our work: to bridge the gap between image quality and non-invasiveness in modern GI diagnostics.

Super-resolution (SR) techniques—especially recent advances using deep generative models—offer promising solutions. Among them, diffusion-based methods such as SR3 (Super-Resolution via Iterative Refinement)~\cite{saharia2021image} have shown outstanding performance in recovering fine-grained details from severely degraded inputs. By treating super-resolution as a conditional generation task, SR3 iteratively denoises a noisy version of the upsampled low-resolution image until convergence to a sharp, high-resolution output.

\begin{figure}[t]
    \centering
    \includegraphics[width=0.7\textwidth]{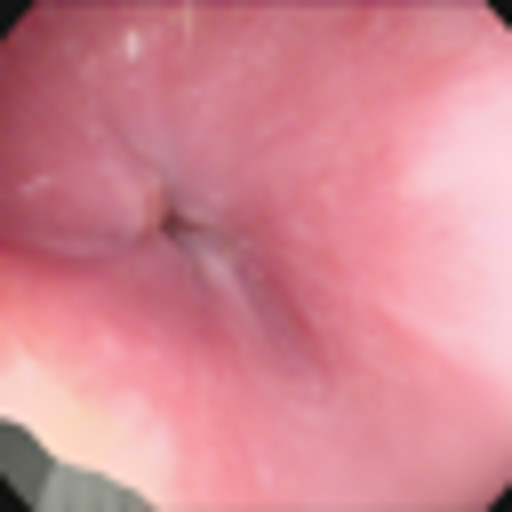}
    \caption{Example of low-resolution capsule endoscopy image showing limited detail for diagnostic purposes.}
    \label{fig:low_res_example}
\end{figure}

The primary objective of this work is to explore and improve diffusion-based super-resolution strategies—starting with SR3 and evolving toward more efficient latent-diffusion approaches—for enhancing capsule endoscopy image resolution. Using the HyperKvasir dataset~\cite{Borgli2020} as a training benchmark, we aim to develop a scalable, semantically-aware resolution enhancement pipeline that could eventually support real-world medical applications.

\subsection{Challenges and Contributions}

Despite the recent success of diffusion models, several challenges persist in the context of medical super-resolution:

\begin{itemize}
    \item \textbf{High computational cost:} Traditional DDPM models such as SR3 operate in pixel space, requiring thousands of sampling steps and high-resolution memory usage.
    \item \textbf{Sensitivity to image noise and clinical variations:} Capsule endoscopy images often suffer from motion blur, low contrast, and lighting inconsistencies.
    \item \textbf{Semantic inconsistency:} Super-resolution models may enhance image sharpness but fail to preserve anatomical correctness if not appropriately trained.
\end{itemize}

To address these issues, our contributions in this work are threefold:

\begin{enumerate}
    \item We implement and optimize a first- and second-generation SR3-based model tailored for capsule endoscopy imagery, including architecture customization and step-wise upscaling strategies.
    \item We conduct comprehensive experiments demonstrating quantitative improvements (PSNR: 27.5 → 29.3 dB, SSIM: 0.65 → 0.71) through architectural enhancements including attention mechanisms and optimized training procedures.
    \item We propose a third-generation model based on latent diffusion by integrating a pretrained VAE from the Stable Diffusion architecture~\cite{rombach2022high}, enabling semantic compression and faster inference in latent space as future work.
\end{enumerate}


\section{Related Work}
\label{sec:related}

This section reviews the existing literature related to diffusion-based image generation and super-resolution, with a particular focus on their applications in the medical imaging domain.

\subsection{Diffusion-Based Super-Resolution}

Diffusion-based models have recently demonstrated significant success in the area of image generation and super-resolution. One notable contribution is SR3~\cite{saharia2021image}, which formulates the super-resolution task as a denoising problem using a score-based generative model. SR3 progressively refines a low-resolution image toward high-resolution through a stochastic denoising process. This approach achieves impressive visual quality, especially in facial and natural image domains.

To address SR3's limitations in terms of computational and memory costs due to pixel-space modeling, Latent Diffusion Models (LDMs)~\cite{rombach2022high} were introduced. These models compress images into a semantic latent space using a pretrained Variational Autoencoder (VAE), allowing diffusion to operate in a more efficient representation. While these methods show great potential, their applications in domain-specific settings such as medical imaging remain relatively underexplored.

\subsection{Medical Image Super-Resolution}

The reviewed literature reveals increasing interest in generative diffusion models for healthcare applications. Compared to GAN-based methods, diffusion models offer more stable training and superior image fidelity, which are essential in clinical scenarios. For instance, Mahapatra et al.~\cite{MAHAPATRA201930} introduced GAN-based models with perceptual loss for retinal and MRI image enhancement. However, such models risk generating anatomically inconsistent hallucinations.

Med-DDPM~\cite{wu2024medddpm} demonstrates that DDPM-based models can be successfully adapted for CT and MR image reconstruction by incorporating domain priors. Likewise, Polyp-DDPM~\cite{wu2023polyp} applies diffusion models to generate structure-preserving polyp images, enhancing segmentation model training. Nevertheless, both works prioritize generation or synthesis rather than resolution enhancement. Moreover, gastrointestinal imaging—especially non-invasive capsule endoscopy—remains under-addressed in terms of domain-specific super-resolution.


\section{Theoretical Background and Motivation}
\label{sec:theory}

In this section, we introduce the theoretical principles behind image super-resolution, with a focus on diffusion probabilistic models.

\subsection{Background on Image Super-Resolution}

Image super-resolution (SR) refers to the process of reconstructing a high-resolution (HR) image from its low-resolution (LR) counterpart. It plays a crucial role in medical imaging applications where high-quality visuals are essential for diagnosis, especially in non-invasive procedures such as capsule endoscopy.

Traditional SR methods include interpolation-based approaches, CNN-based mappings, adversarial GAN models, and more recently, diffusion-based models. In the context of this project, we evaluated multiple architectures before choosing SR3 as the final model.

\subsubsection{Initial Model Consideration: ESRGAN}

In the early stage of the project, ESRGAN (Enhanced Super-Resolution Generative Adversarial Network)~\cite{Wang2018ESRGAN} was considered as the primary candidate. ESRGAN is a widely adopted GAN-based model that enhances perceptual quality by introducing a relativistic discriminator and residual-in-residual blocks.

\begin{figure}[h]
    \centering
    \includegraphics[width=\textwidth]{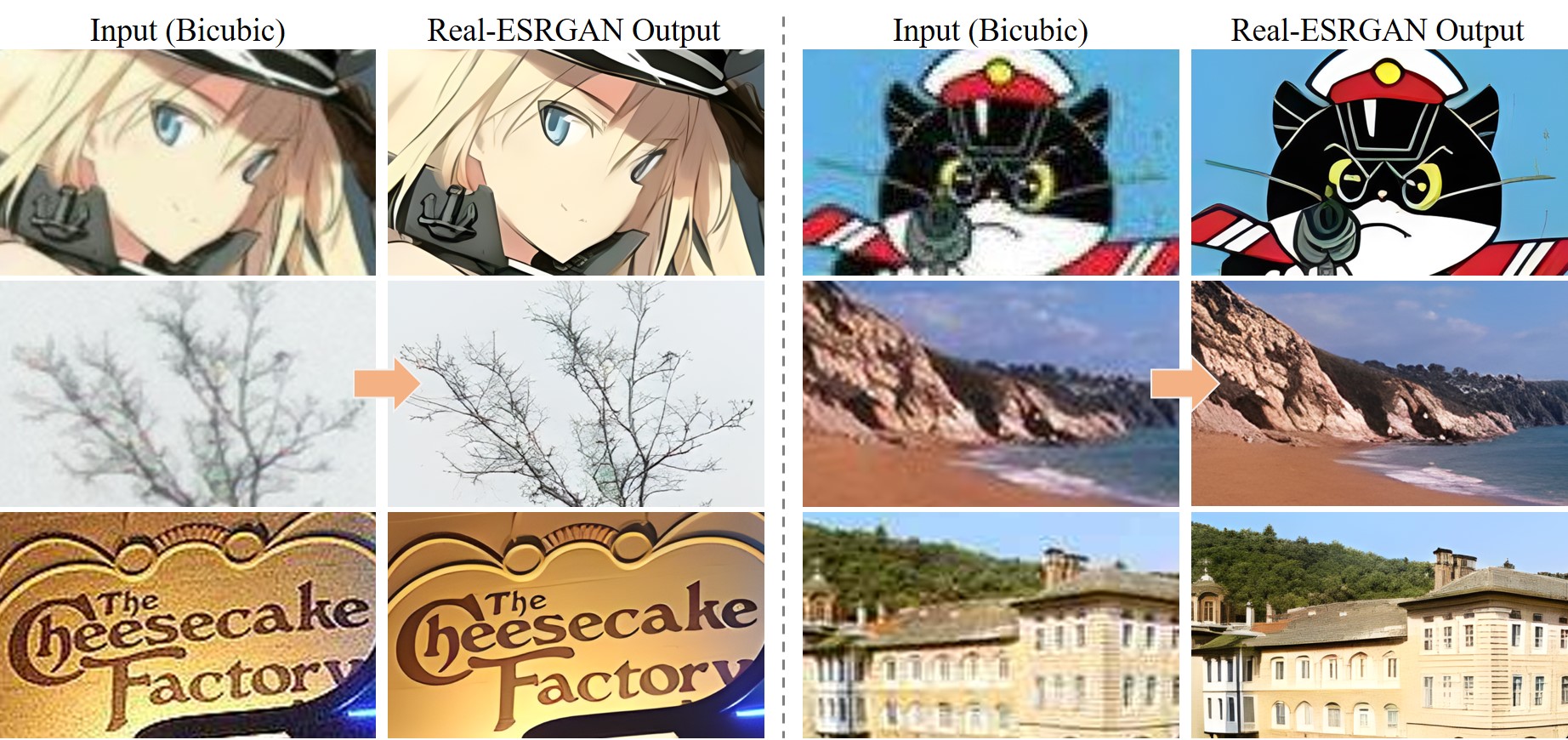}
    \caption{Visual results from Real-ESRGAN: comparing bicubic input (left) and ESRGAN-enhanced output (right).}
    \label{fig:esrgan_visual}
\end{figure}

\begin{figure}[h]
    \centering
    \includegraphics[width=0.9\textwidth]{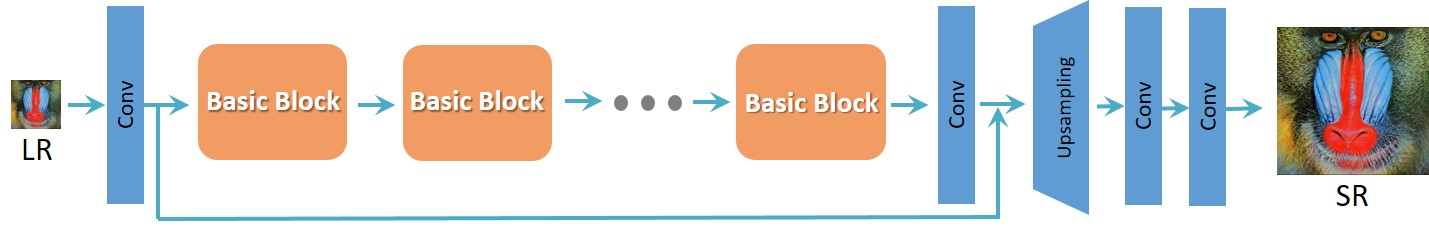}
    \caption{Simplified architecture of the ESRGAN super-resolution network.}
    \label{fig:esrgan_architecture}
\end{figure}

Figure~\ref{fig:esrgan_visual} shows the visual enhancement capability of ESRGAN on various types of images. The architecture, illustrated in Figure~\ref{fig:esrgan_architecture}, is built upon a deep residual-in-residual dense network (RRDB) backbone, with an upsampling module at the output stage.

\subsubsection{Why ESRGAN Was Reconsidered}

Despite its popularity, ESRGAN posed several limitations for our application:
\begin{itemize}
    \item GANs are often unstable to train and require meticulous hyperparameter tuning.
    \item They may introduce hallucinated artifacts, which can be misleading in clinical interpretation.
    \item ESRGAN lacks uncertainty modeling, which is increasingly important in medical domains.
\end{itemize}

\subsubsection{Transition to SR3 Diffusion Model}

To address the above issues, we explored SR3~\cite{saharia2021image}, a diffusion-based generative model. Unlike GANs, SR3 progressively refines noise into high-resolution images via denoising steps, conditioned on low-resolution inputs. Its advantages include:
\begin{itemize}
    \item Improved performance in PSNR and SSIM on structural content
    \item Reduced hallucinated features and enhanced anatomical consistency
    \item More stable training without adversarial loss
\end{itemize}

\subsection{SR3 Model: A Detailed Overview}

The SR3 (Super-Resolution via Iterative Refinement) model, proposed by Saharia et al.~\cite{saharia2021image}, is a diffusion-based super-resolution model designed to generate high-resolution images from low-resolution inputs. This model adapts Denoising Diffusion Probabilistic Models (DDPM) to conditional image generation, achieving super-resolution through a stochastic iterative denoising process.

\begin{figure}[h]
\centering
\includegraphics[width=0.6\textwidth]{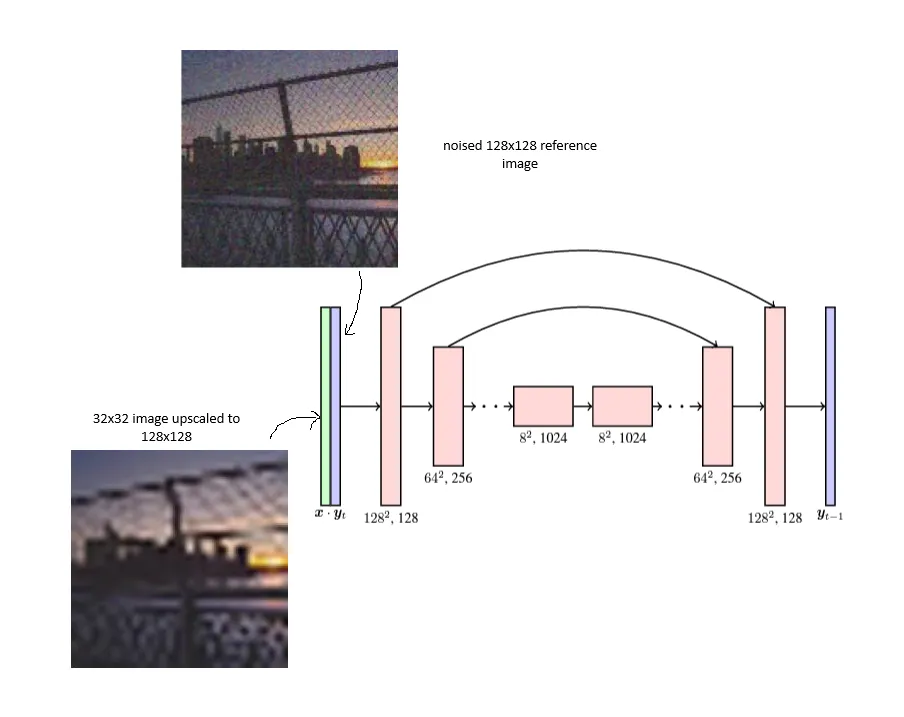}
\caption{SR3 Super-Resolution Process: Starting with low-resolution input, the model refines the image over multiple iterations using learned noise predictions to enhance resolution.}
\label{fig:sr3_process}
\end{figure}

\subsubsection{Forward Diffusion Process}

In SR3, the forward process gradually adds Gaussian noise to the high-resolution image over several time steps. This process can be represented as a Markov chain, where each step involves adding a small amount of noise. Mathematically, the forward diffusion process $q(\mathbf{x}_t|\mathbf{x}_{t-1})$ is defined as:
\begin{equation}
q(\mathbf{x}_{t} | \mathbf{x}_{t-1}) = \mathcal{N}(\mathbf{x}_t; \sqrt{\alpha_t}\mathbf{x}_{t-1}, (1-\alpha_t)\mathbf{I}),
\end{equation}
where $\mathbf{x}_t$ represents the noisy image at step $t$, and $\alpha_t$ controls the variance of the added noise. The process starts with the original high-resolution image $\mathbf{x}_0$ and iteratively adds noise until reaching a pure Gaussian noise image $\mathbf{x}_T$ at step $T$.

\subsubsection{Reverse Process: Iterative Denoising}

The goal of the SR3 model is to learn a reverse process that gradually denoises the image, starting from pure Gaussian noise $\mathbf{x}_T$ and moving back to a clean high-resolution image $\mathbf{x}_0$. The reverse process is parameterized by a neural network (typically a U-Net), which learns to predict the noise added at each step and remove it. The reverse process is defined as:
\begin{equation}
p_\theta(\mathbf{x}_{t-1} | \mathbf{x}_t, \mathbf{z}) = \mathcal{N}(\mathbf{x}_{t-1}; \mu_\theta (\mathbf{x}_t, \mathbf{z}, t), \Sigma_\theta (\mathbf{x}_t, t)),
\end{equation}
where $\mu_\theta$ is the predicted mean by the network, conditioned on both the noisy image $\mathbf{x}_t$ and the low-resolution input $\mathbf{z}$, and $\Sigma_\theta$ represents the predicted variance.

\begin{figure}[h]
\centering
\includegraphics[width=0.8\textwidth]{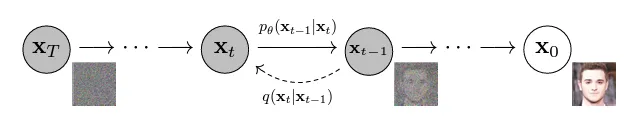}
\caption{SR3 Reverse Steps: Starting with a noisy upsampled input, the model iteratively predicts and removes noise over $T$ timesteps to refine the image and generate a high-resolution output.}
\label{fig:reverse_steps}
\end{figure}

\subsubsection{Training Objective}

The training objective for the SR3 model is based on a denoising objective. The model is trained to predict the added noise $\epsilon$ at each step. The training loss is formulated as:
\begin{equation}
L_\theta = \mathbb{E}_{t,\epsilon}\left[\|\epsilon - \epsilon_\theta(\mathbf{x}_t, \mathbf{z}, t)\|^2\right],
\end{equation}
where $\epsilon_\theta$ is the noise predicted by the model and $\epsilon$ is the true noise added during the forward process. The model minimizes this loss across all time steps, ensuring that it can effectively predict and remove noise at each stage of the reverse process.

\subsection{Future Direction: Latent Diffusion via VAE Compression}

While the pixel-space SR3 model serves as a strong foundation, further efficiency and semantic fidelity can be achieved through latent-space modeling. Inspired by the Stable Diffusion architecture~\cite{rombach2022high}, we propose a third-generation model that introduces a VAE encoder–decoder pipeline into the super-resolution process.

By introducing a pretrained VAE encoder–decoder pair, we aim to:
\begin{itemize}
    \item Compress the image to a semantic latent space (e.g., $64 \times 64 \times 4$)
    \item Perform diffusion in this lower-dimensional latent domain
    \item Decode the denoised latent back to high-resolution pixel space
\end{itemize}

\begin{figure}[h]
    \centering
    \includegraphics[width=\linewidth]{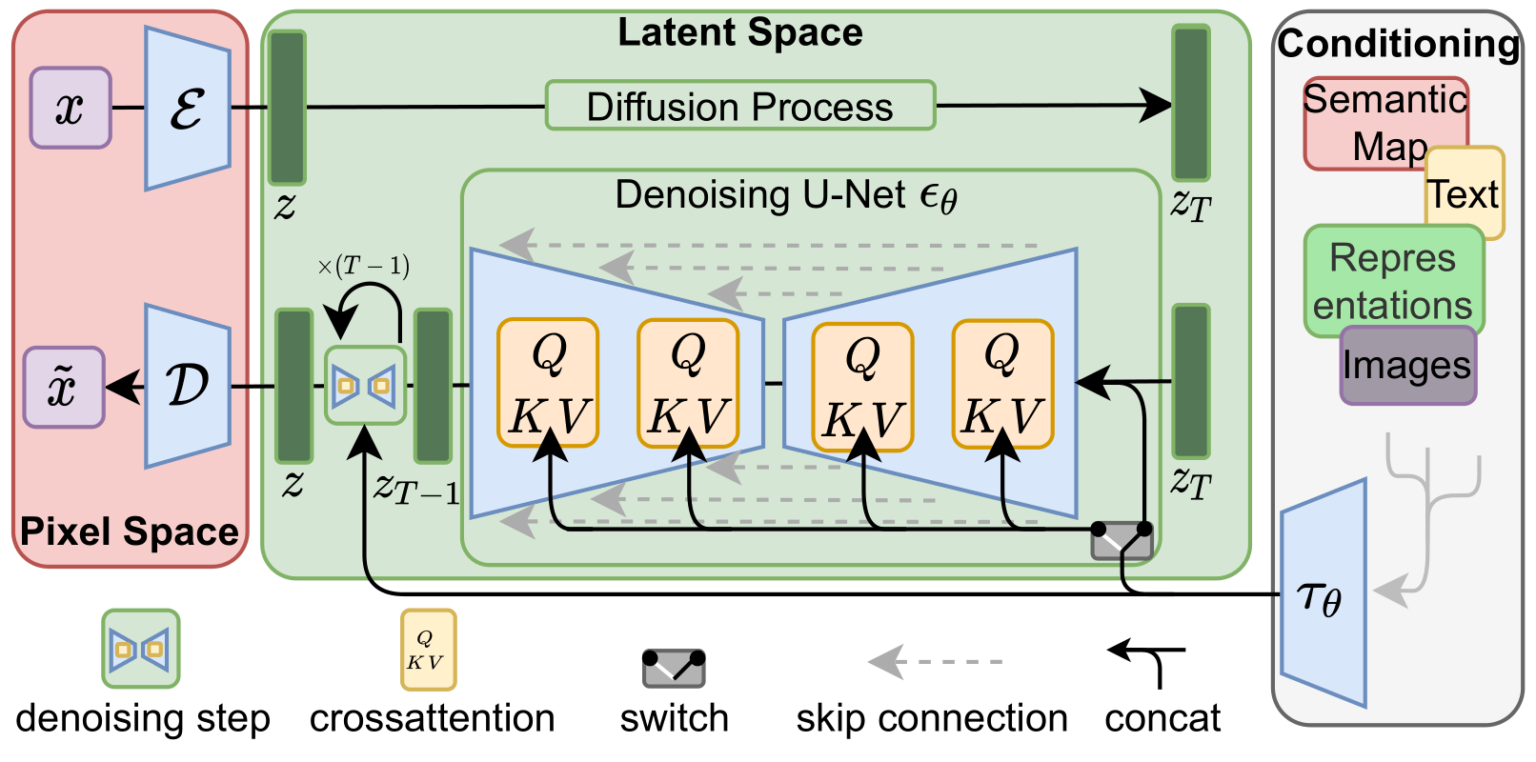}
    \caption{Stable Diffusion-style VAE-enhanced SR architecture. A pretrained encoder maps images to latent space where denoising occurs, and a decoder reconstructs the image~\cite{rombach2022high}.}
    \label{fig:stable_diffusion}
\end{figure}

This approach mirrors the design of Stable Diffusion~\cite{rombach2022high}, enabling efficient modeling of long-range dependencies and reducing training and sampling costs by orders of magnitude.


\section{Methodology}
\label{sec:methodology}

In this section, we present the methodological framework underpinning our approach to high-resolution reconstruction of capsule endoscopy images.

\subsection{Implementation of SR3 Model}

The denoising model implemented in this work follows the general U-Net architecture~\cite{ronneberger2015u}, which is widely regarded as a powerful and flexible backbone for image-to-image translation tasks. The encoder-decoder configuration of U-Net, augmented with symmetric skip connections, ensures that spatial information is effectively propagated throughout the network.

In this study, the model is trained to perform conditioned super-resolution, transforming low-resolution inputs of shape $3 \times 64 \times 64$ into high-resolution outputs of size $3 \times 512 \times 512$, corresponding to an 8× upscaling ratio. Training follows the Denoising Diffusion Probabilistic Model (DDPM) paradigm, in which the clean high-resolution target is gradually perturbed through a forward diffusion process. The training loss is defined as the expected mean squared error between the predicted and true noise:
\[
L_\theta = \mathbb{E}_{t, \epsilon}\left[\|\epsilon - f_\theta(\mathbf{x}_t, \mathbf{z}, t)\|^2\right],
\]
which encourages the model to progressively denoise and restore structural coherence as diffusion steps are reversed.

This implementation employs Group Normalization with 16 groups, Automatic Mixed Precision (AMP) training to reduce GPU memory consumption, and a cosine noise schedule~\cite{nichol2021improved} to modulate the variance of the forward noise injection.

\subsection{Training and Inference Procedures}

To operationalize the SR3 framework in our super-resolution task, we adopt two primary algorithmic phases: a training phase for learning the denoising model, and an inference phase for generating high-resolution outputs via iterative refinement.

\begin{algorithm}
\caption{Training a denoising model $f_{\theta}$}
\begin{algorithmic}[1]
\Repeat
    \State $(x, y_0) \sim p(x, y)$
    \State $\gamma \sim p(\gamma)$
    \State $\epsilon \sim \mathcal{N}(0, I)$
    \State Take a gradient descent step on
    \[
    \nabla_\theta \| f_\theta(x, \sqrt{\gamma}y_0 + \sqrt{1-\gamma}\epsilon, \gamma) - \epsilon \|_p^p
    \]
\Until{converged}
\end{algorithmic}
\end{algorithm}

\begin{algorithm}
\caption{Inference in $T$ iterative refinement steps}
\begin{algorithmic}[1]
\State $y_T \sim \mathcal{N}(0, I)$
\For{$t = T, \dots, 1$}
    \If{$t > 1$}
        \State $z \sim \mathcal{N}(0, I)$
    \Else
        \State $z = 0$
    \EndIf
    \State $y_{t-1} = \frac{1}{\sqrt{\alpha_t}} \left( y_t - \frac{1 - \alpha_t}{\sqrt{1 - \gamma_t}} f_\theta(x, y_t, \gamma_t) \right) + \sqrt{1 - \alpha_t}z$
\EndFor
\State \Return $y_0$
\end{algorithmic}
\end{algorithm}

\subsection{Evaluation Metrics for Image Quality}

To quantitatively assess the performance of image super-resolution models like SR3, two widely used metrics are employed: Peak Signal-to-Noise Ratio (PSNR) and Structural Similarity Index Measure (SSIM).

\subsubsection{Peak Signal-to-Noise Ratio (PSNR)}

PSNR measures the pixel-wise fidelity between the reconstructed image and the ground-truth image. It is defined in decibels (dB) as:
\begin{equation}
\text{PSNR} = 10 \cdot \log_{10} \left( \frac{{\text{MAX}^2}}{{\text{MSE}}} \right),
\end{equation}
where MAX denotes the maximum possible pixel value (usually 255 for 8-bit images), and MSE is the mean squared error between the original and reconstructed images. Higher PSNR values indicate better image fidelity.

\subsubsection{Structural Similarity Index Measure (SSIM)}

SSIM is a perceptual metric that evaluates the structural similarity between two images. It considers luminance, contrast, and structural information~\cite{wang2004image}. The SSIM index between images $x$ and $y$ is computed as:
\begin{equation}
\text{SSIM}(x, y) = \frac{(2\mu_x\mu_y + C_1)(2\sigma_{xy} + C_2)}{(\mu_x^2 + \mu_y^2 + C_1)(\sigma_x^2 + \sigma_y^2 + C_2)},
\end{equation}
where $\mu_x, \mu_y$ are local means, $\sigma_x^2, \sigma_y^2$ are variances, $\sigma_{xy}$ is the covariance, and $C_1, C_2$ are stabilization constants. SSIM ranges from $-1$ to $1$, with $1$ indicating perfect structural similarity.


\section{Experiments and Results}
\label{sec:experiments}

This section presents our experimental setup, dataset description, training procedures, and quantitative/qualitative results.

\subsection{Experimental Setup}

The training and inference experiments were conducted across two distinct computing environments. Table~\ref{tab:env_comparison} summarizes the hardware and software configurations used for both generations of the SR3 model.

\begin{table}[h]
\centering
\caption{Hardware and software configuration for first and second-generation SR3 experiments.}
\label{tab:env_comparison}
\begin{tabular}{lll}
\toprule
\textbf{Component} & \textbf{First-Gen} & \textbf{Second-Gen} \\
\midrule
CPU & Intel Xeon Platinum 8352V & AMD Ryzen 9 7950X \\
GPU & NVIDIA RTX 4090 & NVIDIA RTX 4090 \\
GPU Memory & 24 GB & 48 GB \\
System Memory & 120 GB & 32 GB \\
Operating System & Ubuntu 22.04 LTS & Ubuntu 24.04.2 \\
Python & 3.12 & 3.12 \\
PyTorch & 2.3.0 & 2.3.0 \\
CUDA Toolkit & $\leq$ 12.4 & $\leq$ 12.4 \\
\bottomrule
\end{tabular}
\end{table}

\subsection{Dataset: HyperKvasir}

The HyperKvasir dataset~\cite{Borgli2020} is a large-scale, publicly available collection of gastrointestinal endoscopy images with anatomical and pathological labels. It contains over 110,000 images spanning multiple GI tract regions and pathological conditions. In this project, we use 10,662 labeled images stored in JPEG format, categorized into 23 different classes based on anatomical landmarks, mucosal quality, pathological findings, and therapeutic interventions.

\begin{figure}[h]
\centering
\includegraphics[width=\textwidth]{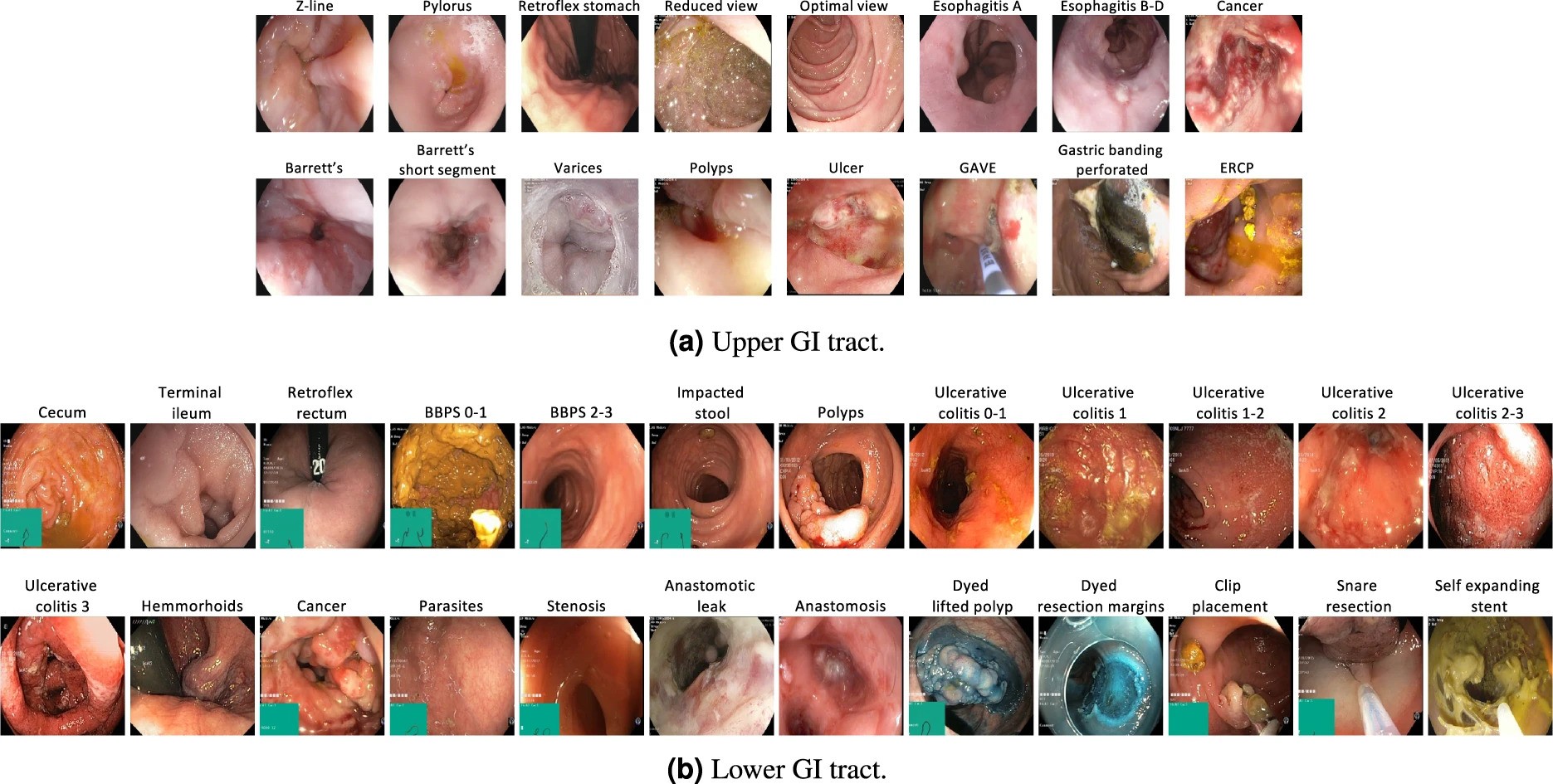}
\caption{Image examples of the various labeled classes for images and/or videos from the HyperKvasir dataset.}
\label{fig:hyperkvasir_samples}
\end{figure}

\begin{figure}[h]
\centering
\includegraphics[width=\textwidth]{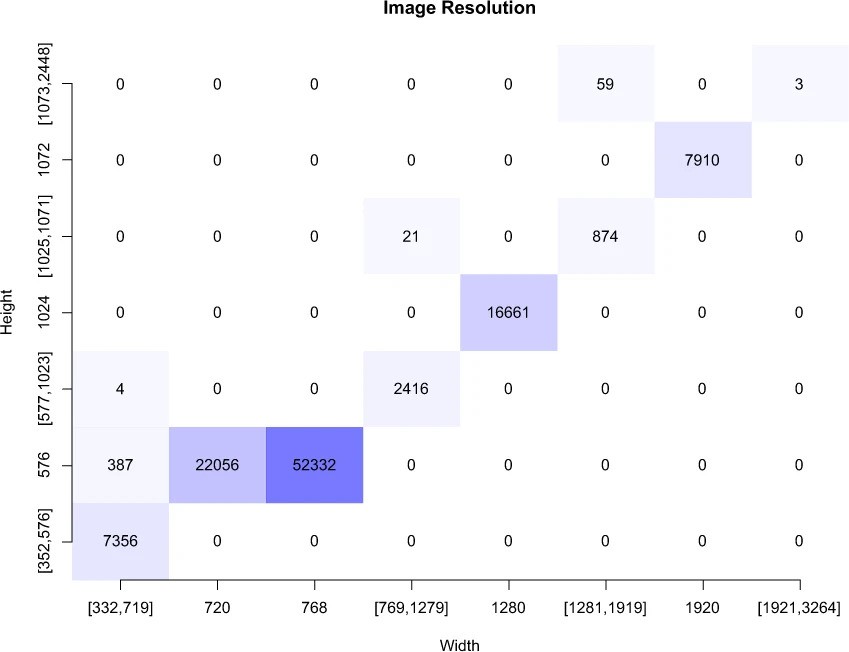}
\caption{Resolution distribution of the 110,079 images in HyperKvasir dataset.}
\label{fig:resolution_dist}
\end{figure}

\begin{figure}[h]
\centering
\includegraphics[width=\textwidth]{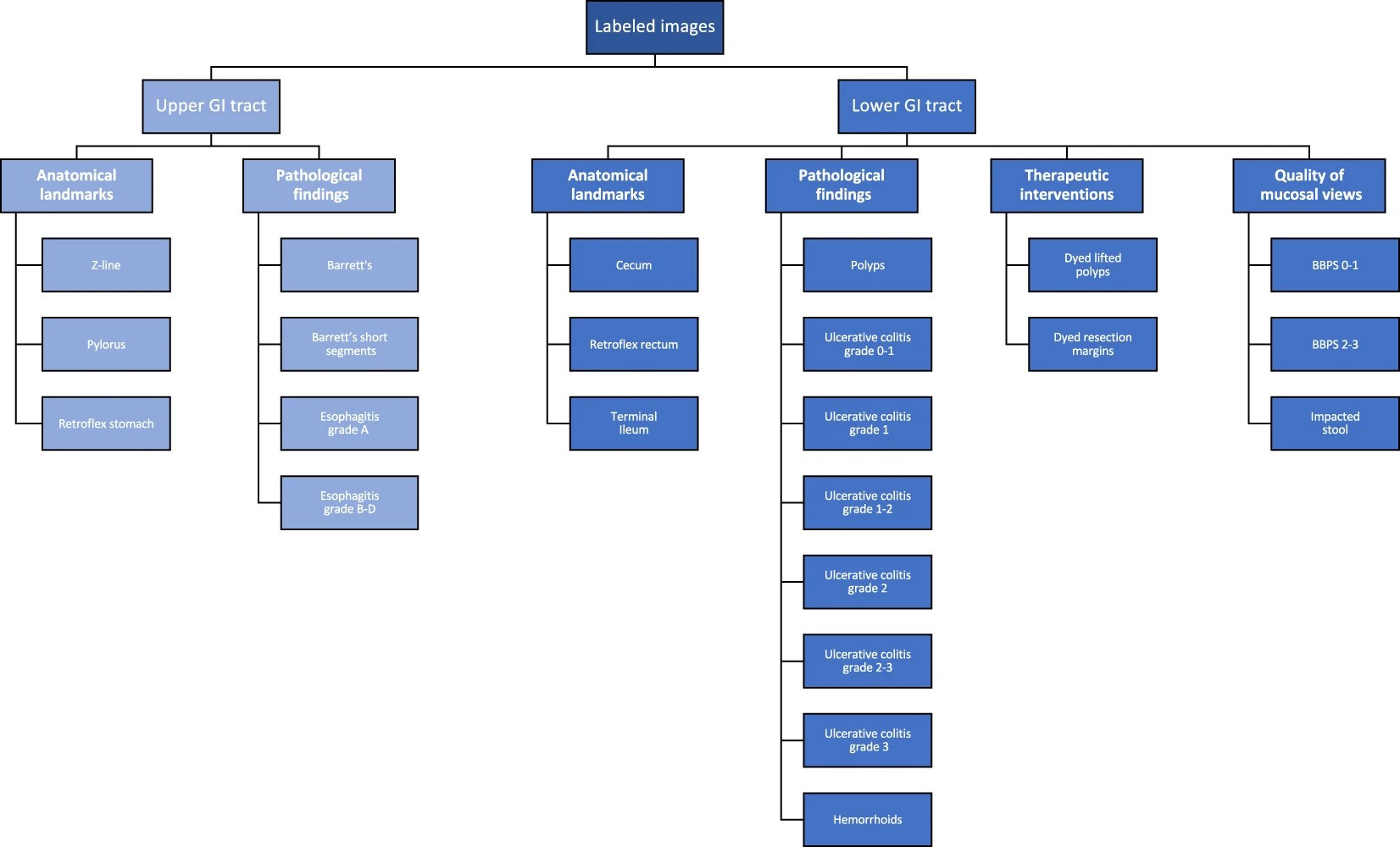}
\caption{The various image classes structured under position and type, showing the organization of the stored images.}
\label{fig:dataset_structure}
\end{figure}

The dataset is organized in a hierarchical folder structure as shown in Figure~\ref{fig:dataset_structure}, with categorization by anatomical location and type of finding.

\subsection{Data Preprocessing}

\begin{figure}[h]
\centering
\includegraphics[width=0.6\textwidth]{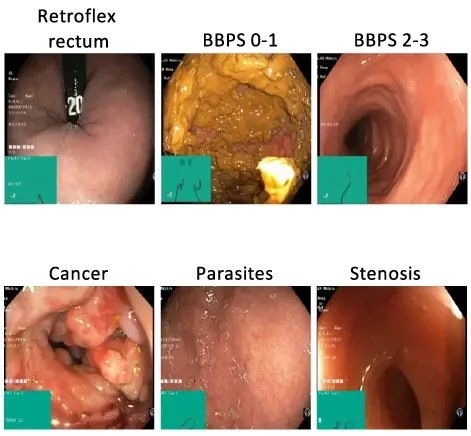}
\caption{Example of a raw image from the dataset with a green block in the bottom-left corner that requires removal.}
\label{fig:dirty_image}
\end{figure}

Before training, we applied several preprocessing steps to clean the dataset. Some raw images contain a green annotation block in the bottom-left corner (Figure~\ref{fig:dirty_image}), which needed to be removed to prevent contamination of the training data.

The preprocessing pipeline included:
\begin{itemize}
    \item Removal of green annotation blocks using HSV color space thresholding
    \item Normalization to [0, 1] range
    \item Random horizontal and vertical flips for data augmentation
    \item Creation of low-resolution inputs via bicubic downsampling from $512 \times 512$ to $64 \times 64$
\end{itemize}

\subsection{Exploratory Data Analysis}

To understand the distributional properties and visual characteristics of the dataset, we conducted comprehensive exploratory data analysis (EDA) on 100 randomly selected image triplets: low-resolution ($64 \times 64$), super-resolved ($512 \times 512$), and high-resolution ground truth ($512 \times 512$).

\begin{figure}[h]
    \centering
    \includegraphics[width=\linewidth]{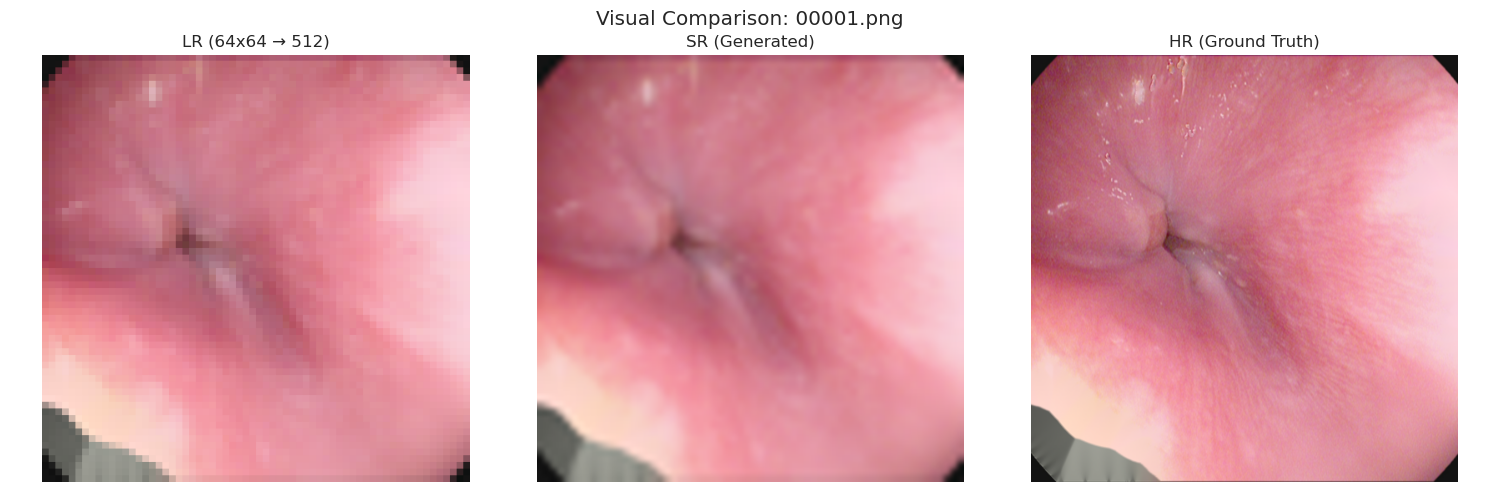}
    \caption{Visual comparison of a low-resolution input image (left), SR3-generated super-resolved output (middle), and ground truth high-resolution image (right).}
    \label{fig:sample_comparison}
\end{figure}

Figure~\ref{fig:sample_comparison} presents a visual comparison showing the SR image exhibits significant perceptual improvements over the LR input and retains anatomical structures consistent with the HR reference.

\begin{figure}[h]
    \centering
    \includegraphics[width=0.7\linewidth]{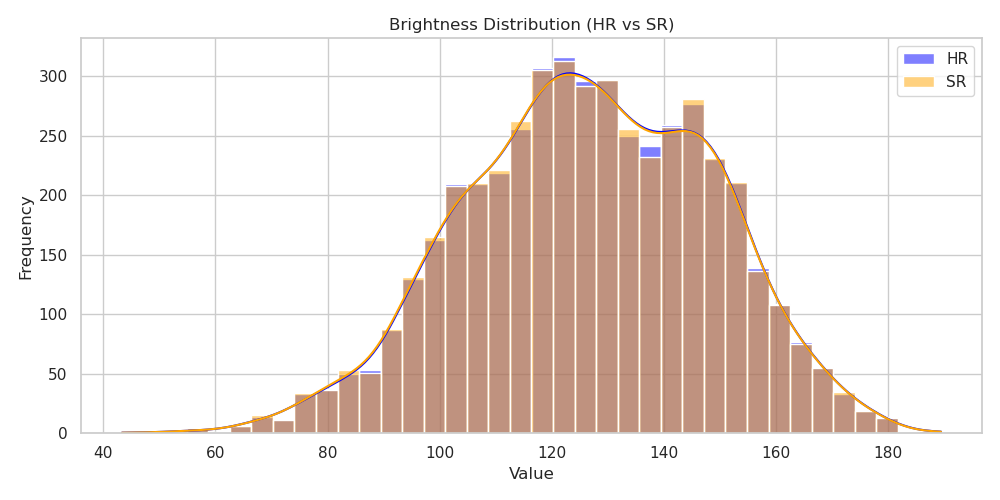}
    \caption{Histogram distribution of grayscale brightness values for SR and HR images showing close alignment.}
    \label{fig:brightness_dist}
\end{figure}

\begin{figure}[h]
    \centering
    \includegraphics[width=0.7\linewidth]{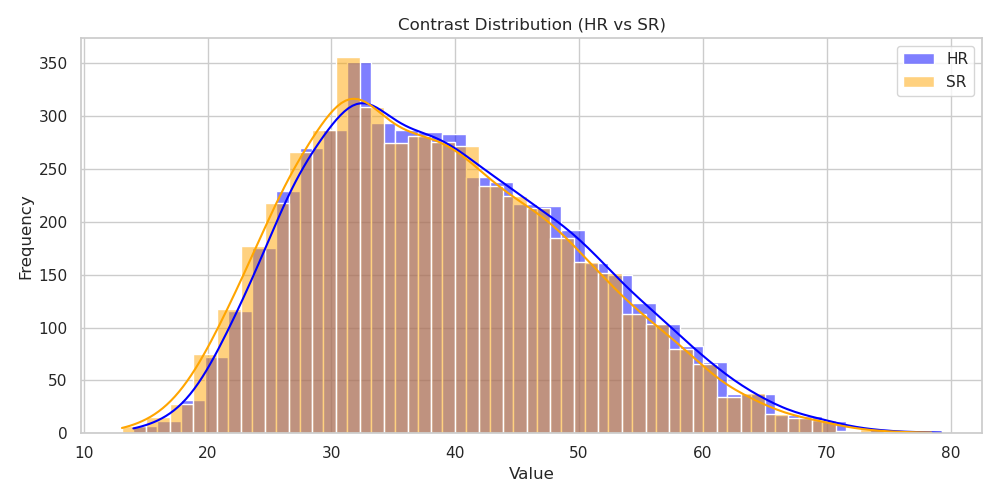}
    \caption{Histogram distribution of grayscale contrast (standard deviation) for SR and HR images.}
    \label{fig:contrast_dist}
\end{figure}

Statistical analysis revealed close alignment between SR and HR images in terms of brightness (Figure~\ref{fig:brightness_dist}) and contrast (Figure~\ref{fig:contrast_dist}) distributions.

The average PSNR between SR and HR images was measured at \textbf{19.2 dB}, and the mean SSIM was \textbf{0.79}, indicating moderate-to-high level of perceptual similarity crucial for diagnostic medical imaging.

\subsection{Training Configuration}

\subsubsection{First-Generation Model}

The first-generation SR3 model was trained with the following configuration:
\begin{itemize}
    \item U-Net with 5 resolution levels and channel multipliers [1, 2, 4, 8, 16]
    \item Single residual block per resolution level
    \item 2000-step linear beta schedule ($\beta_{\text{start}} = 10^{-6}$, $\beta_{\text{end}} = 10^{-2}$)
    \item Batch size: 8
    \item Learning rate: $3 \times 10^{-6}$ with Adam optimizer
    \item Training precision: FP32
    \item Total iterations: 1,000,000
    \item Validation frequency: every 10,000 iterations
\end{itemize}

\subsubsection{Second-Generation Model}

The second-generation model incorporated the following enhancements:
\begin{itemize}
    \item Two residual blocks per resolution level (increased depth)
    \item Self-attention modules at 16×16, 32×32, and 64×64 resolutions
    \item Cosine noise scheduling~\cite{nichol2021improved} (more stable than linear)
    \item Dropout regularization ($p = 0.1$) after second convolution in each block
    \item FP16 mixed-precision training with AMP
    \item Exponential Moving Average (EMA) of weights with decay 0.9999
    \item Gradient clipping with max L2 norm of 1.0
    \item Learning rate decay at iterations [150k, 230k] with $\gamma = 0.5$
\end{itemize}

\subsection{Quantitative Results}

Table~\ref{tab:results_comparison} summarizes the quantitative performance comparison between bicubic interpolation and the two generations of SR3 models.

\begin{table}[h]
\centering
\caption{Quantitative comparison of super-resolution methods on HyperKvasir validation set.}
\label{tab:results_comparison}
\begin{tabular}{lcc}
\toprule
\textbf{Method} & \textbf{PSNR (dB)} & \textbf{SSIM} \\
\midrule
Bicubic Interpolation & 24.3 & 0.58 \\
First-Gen SR3 (Baseline) & 27.5 & 0.65 \\
Second-Gen SR3 (+ Attention) & \textbf{29.3} & \textbf{0.71} \\
\bottomrule
\end{tabular}
\end{table}

As shown, the second-generation SR3 model achieves significant improvements:
\begin{itemize}
    \item PSNR improvement: 24.3 dB (bicubic) → 27.5 dB (Gen-1) → 29.3 dB (Gen-2)
    \item SSIM improvement: 0.58 (bicubic) → 0.65 (Gen-1) → 0.71 (Gen-2)
    \item Inference time: reduced from ~3-4 seconds to ~2 seconds per image
\end{itemize}

\begin{figure}[h]
    \centering
    \begin{subfigure}[b]{0.48\textwidth}
        \includegraphics[width=\textwidth]{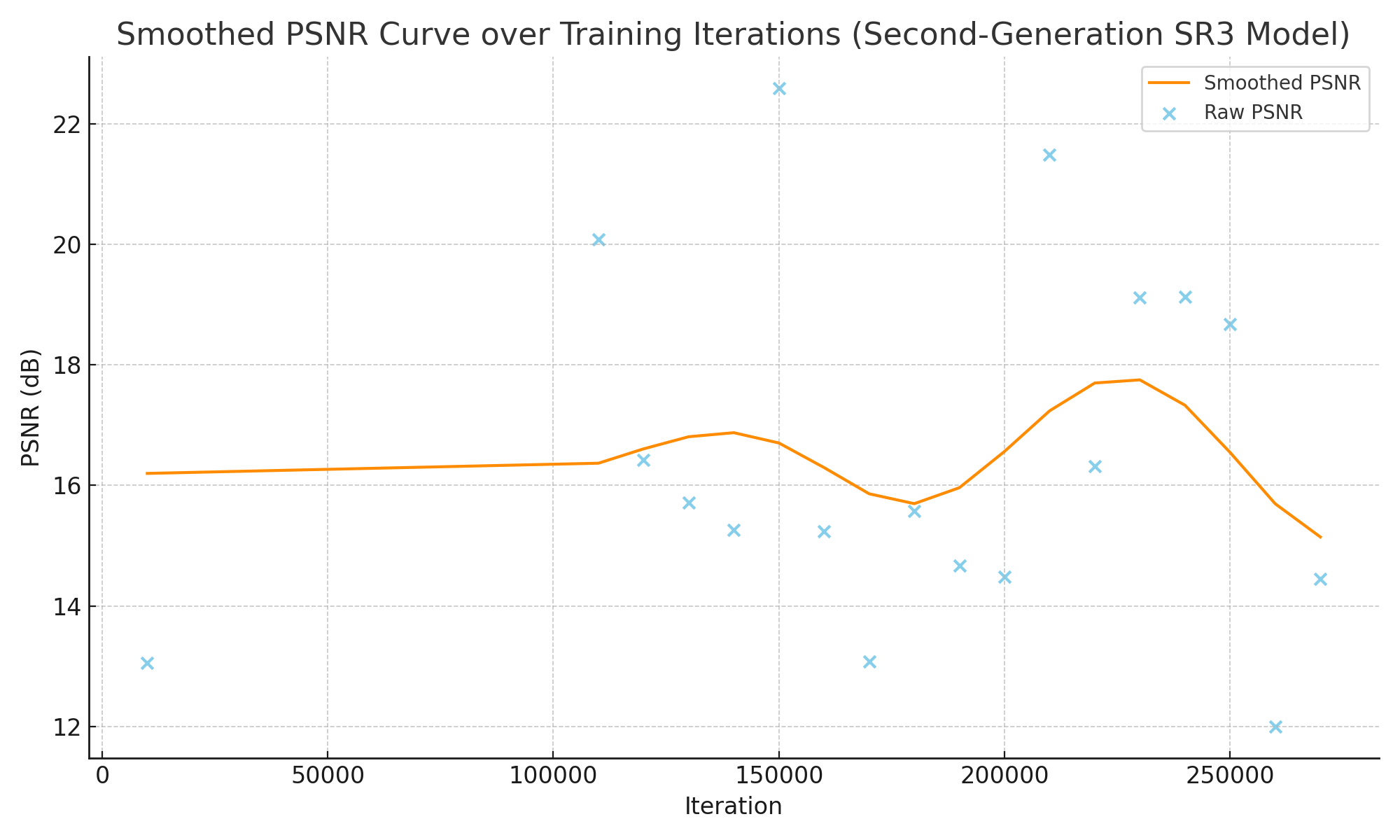}
        \caption{First-generation training curve}
    \end{subfigure}
    \hfill
    \begin{subfigure}[b]{0.48\textwidth}
        \includegraphics[width=\textwidth]{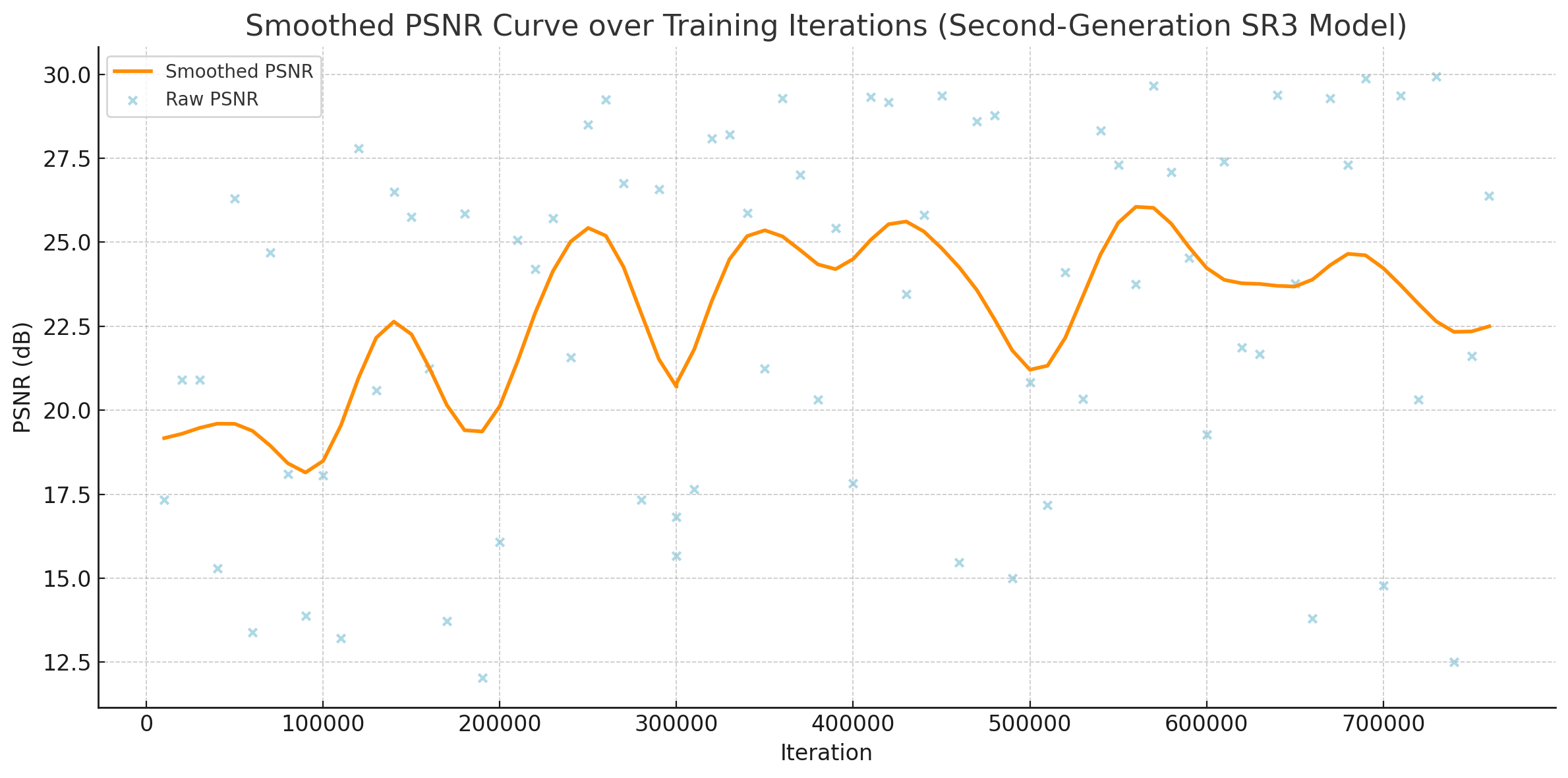}
        \caption{Second-generation training curve (smoothed)}
    \end{subfigure}
    \caption{PSNR training curves comparing first and second-generation SR3 models.}
    \label{fig:training_curves}
\end{figure}

Figure~\ref{fig:training_curves} shows the PSNR training curves for both generations, demonstrating smoother convergence for the second-generation model with cosine scheduling.

\subsection{Qualitative Analysis}

\begin{figure}[h]
    \centering
    \begin{subfigure}[b]{0.48\textwidth}
        \includegraphics[width=\textwidth]{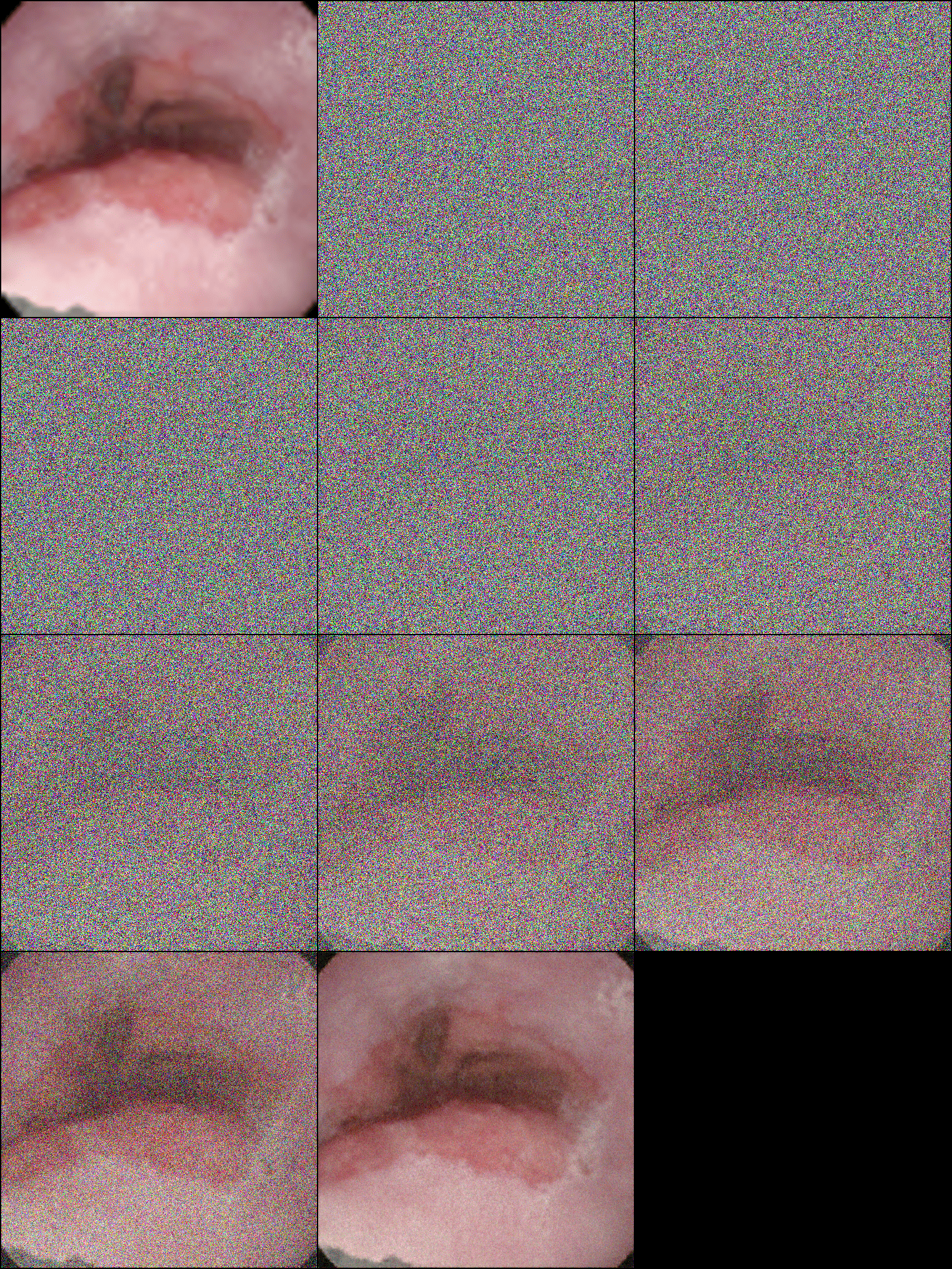}
        \caption{First-generation validation sample 1}
    \end{subfigure}
    \hfill
    \begin{subfigure}[b]{0.48\textwidth}
        \includegraphics[width=\textwidth]{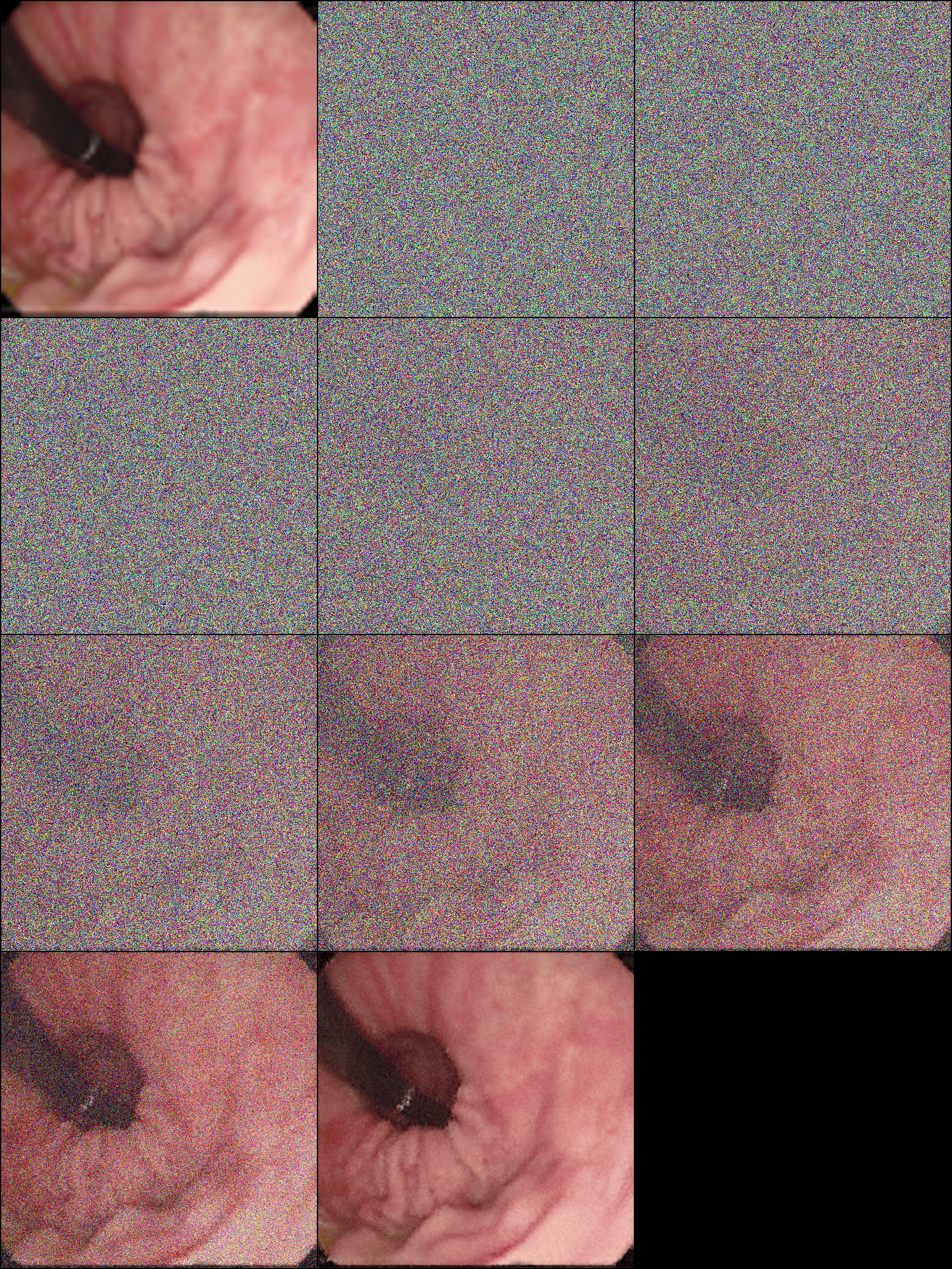}
        \caption{First-generation validation sample 2}
    \end{subfigure}
    \caption{Validation examples from first-generation SR3 model.}
    \label{fig:validation_gen1}
\end{figure}

\begin{figure}[h]
    \centering
    \begin{subfigure}[b]{0.48\textwidth}
        \includegraphics[width=\textwidth]{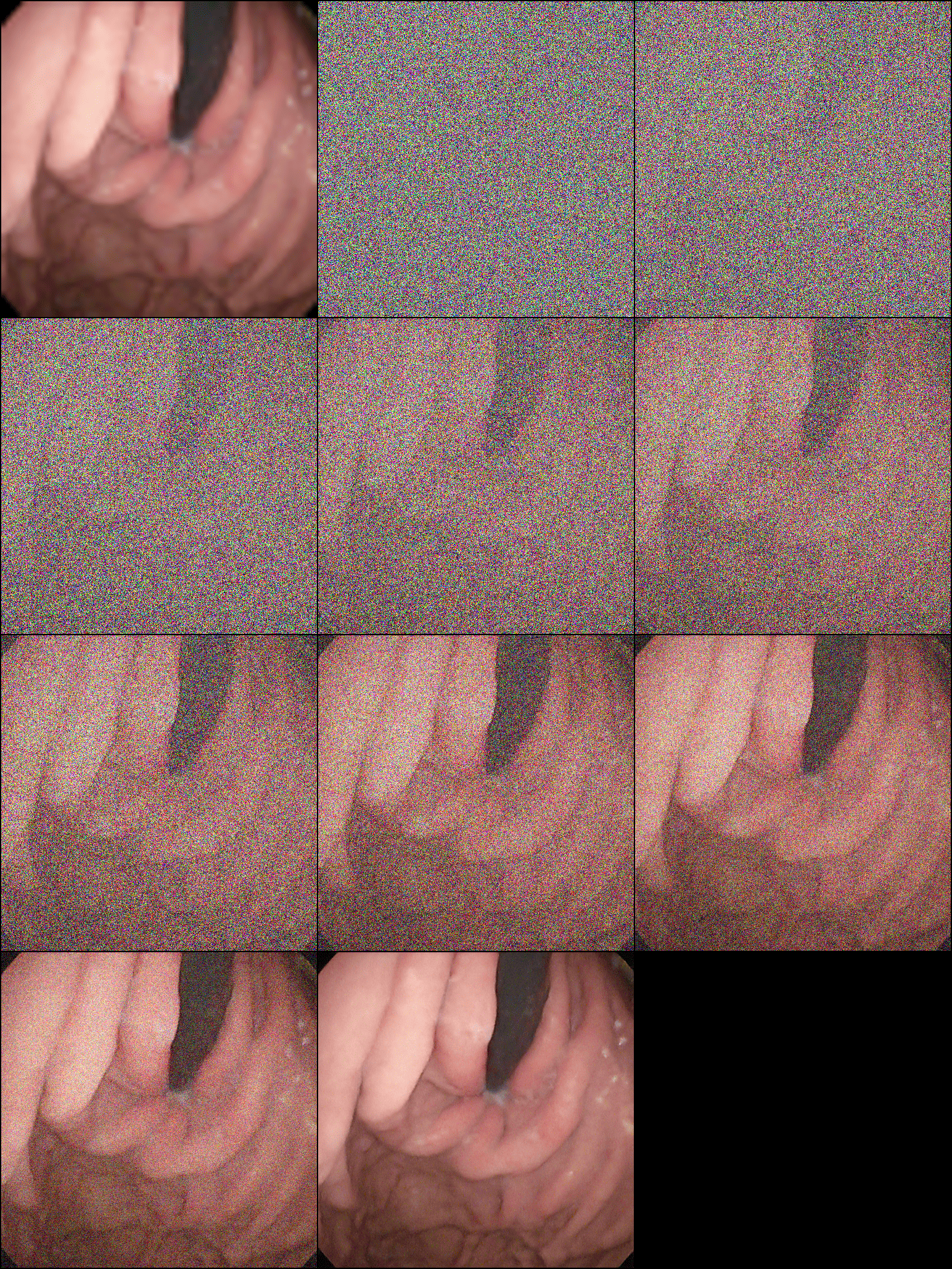}
        \caption{Second-generation validation sample 1}
    \end{subfigure}
    \hfill
    \begin{subfigure}[b]{0.48\textwidth}
        \includegraphics[width=\textwidth]{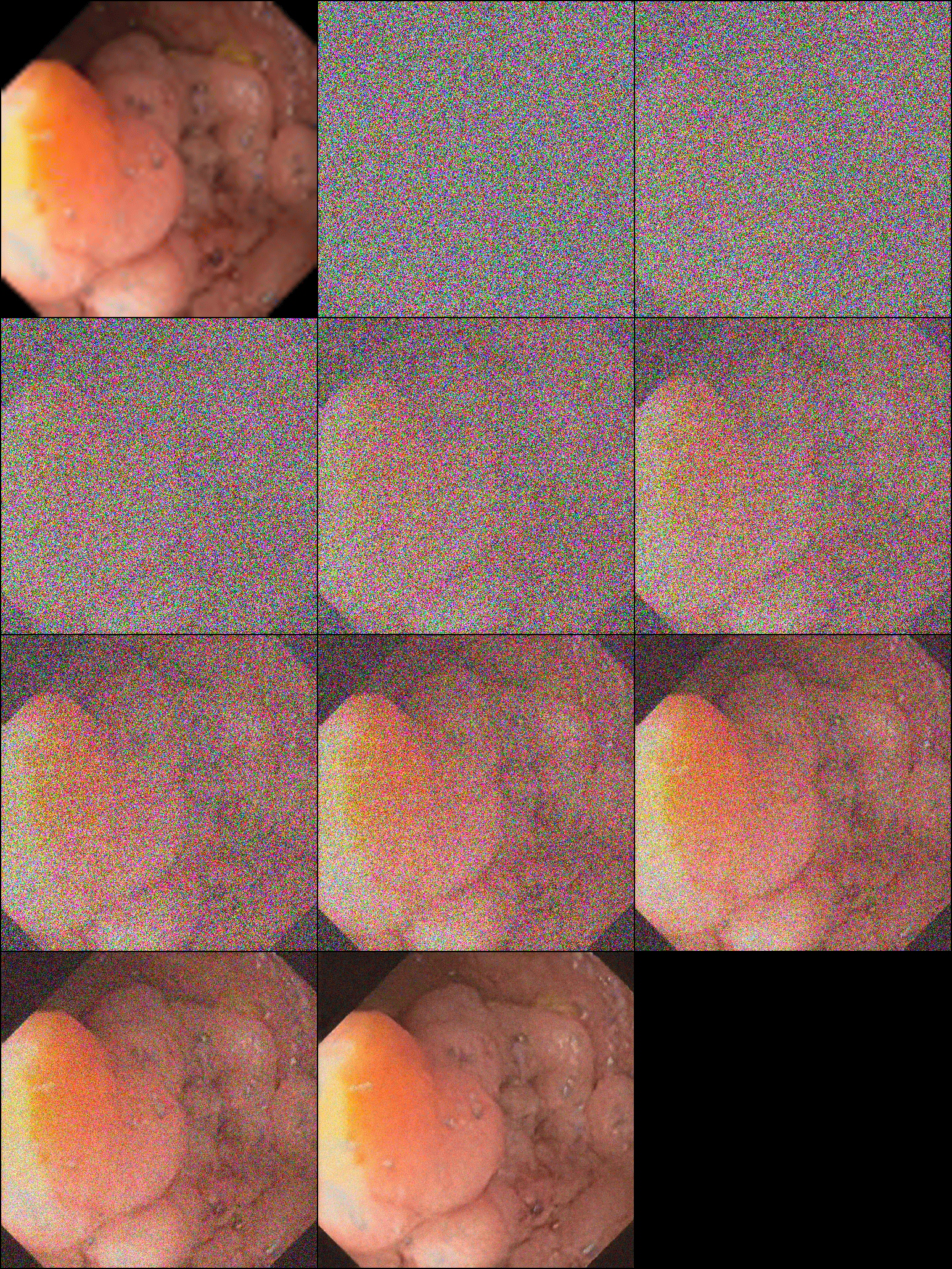}
        \caption{Second-generation validation sample 2}
    \end{subfigure}
    \caption{Validation examples from second-generation SR3 model showing improved detail preservation.}
    \label{fig:validation_gen2}
\end{figure}

Visual inspection of super-resolved images reveals that the SR3 models effectively preserve anatomical details such as mucosal boundaries, vascular patterns, and lesion structures. Figures~\ref{fig:validation_gen1} and~\ref{fig:validation_gen2} present validation examples from both model generations.

Unlike GAN-based methods such as ESRGAN, the diffusion-based approach avoids introducing unrealistic artifacts or hallucinations that could mislead clinical interpretation. The attention mechanism in the second generation further enhances the preservation of fine-grained structures.

\subsection{Model Comparison}

Table~\ref{tab:model_comparison} provides a comprehensive comparison between the two generations of SR3 models.

\begin{table}[h]
\centering
\caption{Comprehensive comparison between first and second-generation SR3 models.}
\label{tab:model_comparison}
\resizebox{\textwidth}{!}{
\begin{tabular}{lll}
\toprule
\textbf{Feature} & \textbf{First-Gen SR3} & \textbf{Second-Gen SR3} \\
\midrule
Diffusion Space & Pixel space & Pixel space \\
Guidance Input & Bicubic + HR Noise & Bicubic + HR Noise \\
U-Net Backbone & Vanilla U-Net & U-Net + Self-Attention \\
Residual Blocks & 1 per resolution level & 2 per resolution level \\
Dropout & None & 0.1 \\
Normalization & GroupNorm (16) & GroupNorm (16) \\
Noise Scheduler & Linear Beta (2000 steps) & Cosine Beta (2000 steps) \\
Training Precision & FP32 & FP16 (AMP-enabled) \\
PSNR / SSIM & 27.5 dB / 0.65 & 29.3 dB / 0.71 \\
Inference Time & 3–4 seconds & 2 seconds \\
Notable Benefits & Baseline super-resolution & Higher accuracy, faster \\
\bottomrule
\end{tabular}
}
\end{table}


\section{Conclusion and Future Work}
\label{sec:conclusion}

This work explored the design, implementation, and iterative enhancement of a super-resolution framework tailored for medical imaging—specifically gastrointestinal endoscopy based on diffusion probabilistic models.

\subsection{Summary of Contributions}

Our first-generation model established a strong foundation by implementing the original SR3 framework with a vanilla U-Net architecture, achieving PSNR of 27.5 dB and SSIM of 0.65. The second-generation model introduced architectural and optimization advancements including:
\begin{itemize}
    \item Self-attention mechanisms at multiple resolutions
    \item Increased residual depth (2 blocks per level)
    \item Cosine noise scheduling for smoother training
    \item FP16 mixed-precision training
    \item Dropout regularization and gradient clipping
\end{itemize}

These improvements resulted in performance gains to PSNR of 29.3 dB and SSIM of 0.71, along with reduced inference time and better anatomical structure preservation.

\subsection{Future Directions}

Despite improvements, pixel-space modeling presents computational limitations. Our future work will focus on:

\begin{enumerate}
    \item \textbf{Latent Diffusion Modeling:} Using VAE-based compression (as shown in Figure~\ref{fig:stable_diffusion}) to reduce computational burden while preserving semantic fidelity. This third-generation approach is expected to provide:
    \begin{itemize}
        \item 4–8× faster sampling due to smaller tensor size in latent space
        \item Better semantic consistency through VAE compression
        \item Lower VRAM footprint enabling higher batch sizes
    \end{itemize}

    \item \textbf{Real-Time Inference:} Model compression and optimization for deployment on edge devices using frameworks like NCNN or TensorRT for mobile medical equipment.

    \item \textbf{Clinical Validation:} Collaboration with medical professionals to validate diagnostic utility and ensure anatomical correctness of super-resolved images.

    \item \textbf{Extended Evaluation:} Additional perceptual metrics such as LPIPS (Learned Perceptual Image Patch Similarity) and FID (Fréchet Inception Distance).

    \item \textbf{Deployment Pipeline:} Containerization with Docker, orchestration via Kubernetes, and compliance with medical data standards (DICOM) and privacy regulations (HIPAA, GDPR).
\end{enumerate}

\subsection{Broader Impact}

This work contributes a robust and extensible pipeline for diffusion-based super-resolution in the medical imaging domain. By bridging the resolution gap through data-driven post-processing, the proposed method provides clinicians with clearer visual cues without the need for invasive high-end optical systems. This has practical implications in:
\begin{itemize}
    \item Early diagnosis of gastrointestinal pathologies
    \item Remote screening in resource-constrained environments
    \item Longitudinal patient monitoring with non-invasive procedures
\end{itemize}

In conclusion, the progressive development from baseline SR3 to attention-enhanced models, and the proposed future extension to latent diffusion, sets a foundation for intelligent, data-driven medical imaging systems that can make high-quality diagnostic imaging both accessible and clinically reliable.


\section*{Acknowledgments}

The author thanks Professor Subrota Kumar Mondal for supervision and guidance throughout this research. We acknowledge the HyperKvasir dataset creators for making their data publicly available for research purposes.


\bibliographystyle{unsrtnat}
\bibliography{references}

\end{document}